\documentclass[aps,superscriptaddress,altaffilletter,tightenlines]{revtex4}

\newcommand{\be}{\begin{equation}}
\newcommand{\n}{\label}
\newcommand{\ee}{\end{equation}}
\newcommand{\no}{\noindent}
\newcommand{\ben}{\begin{eqnarray}}
\newcommand{\een}{\end{eqnarray}}

\begin{document}

\author{
 L. P. Chimento$^{2,}\,$$^3$\thanks{
e-mail: chimento@df.uba.ar}, F. Pennini$^{1}$\thanks{
e-mail: pennini@venus.fisica.unlp.edu.ar} and A. Plastino$^{1,}\,$$^3$
\thanks{e-mail: plastino@venus.fisica.unlp.edu.ar}}
\affiliation{$^1$Physics Department, National University La Plata, \\
C. C. 727, 1900 La Plata, Argentina \\
 $^2$Departamento de F\'{\i}sica, Facultad de Ciencias\\
Exactas y Naturales, Universidad de Buenos Aires, Ciudad\\
Universitaria, Pabell\'{o}n I, 1428 Buenos Aires, Argentina \\ 
$^3$Argentine National Research Council (CONICET)}

\title{The Frieden-Soffer Extreme Physical Information Principle in a \\
Non-extensive Setting}

\begin{abstract}

We show that the  Frieden  
and Soffer's  Extreme Physical Information principle, applied to 
 a  non-extensive   
{\it statistical} scenario, yields     
solutions to several  well-known classical {\it dynamical} 
problems.  
\end{abstract}

\maketitle

\section{ Introduction}

\subsection{Fisher's Information for Translation Families}

A very active area of theoretical endeavour nowadays is that concerned
with the investigation of properties and applications of Fisher's
information measure for translation families   (to be denoted
herefrom as $I$), whose applications to diverse problems in theoretical
physics have been pioneered by  Frieden,
Soffer, Nikolov, Silver, 
and others \cite{f1,f2,f3,f4,f5,f6,f7,f8,f9,f10,si93,roy,fnew}. 
They have
unveiled many  properties of Fisher's information measure 
and clarified its relation to Shannon's
logarithmic information measure 

\begin{equation}
S\,=\,-\,\int dx\, f(x) ln f(x) ,  \label{entrop1}
\end{equation}

\noindent
where $f(x)$ is, of course, a probability density (for the sake of
simplicity we restrict our considerations here to the one dimensional
case). 

A word of caution: the information $I$ that we are here considering is that
{\it particular version of the one originally introduced by Fisher} \cite
{f1} that applies for translation families, i.e., refers to a
measure of the  uncertainty in determining {\it a position parameter $
x$} by a maximum likelihood estimation \cite{si93}. It is seen \cite
{f1} that the  best possible estimator of $x$, after a large
number of samples is examined, suffers a mean-square deviation $e^2$ from $
\langle x \rangle $ that minimizes the so-called Cramer-Rao bound \cite
{f1,f2,f3,f4,f5}

\begin{equation}
I\,e^2\,\ge \,1, 
\end{equation}
where Fisher's information $I$ for translation families reads  
\begin{equation}
I\,=\,\int \,dx\,f(x)\,\{ \frac{\frac{df}{dx}}{f(x)} \}^2,
\end{equation}

\noindent
i.e.,

\begin{equation}
I\,=\,< \,\{ \frac{\frac{df}{dx}}{f(x)} \}^2\,> .  \label{ifisher}
\end{equation}

\subsection{Non-extensive Thermostatistics}

Another area of current interest, for a variety of physical reasons, is
that related 
to nonlinear formalisms. Among them we can single out nonextensive
thermostatistics (NET) \cite{t1}, characterized by a (real) 
 parameter $q$.   NET refers, of course, to the use of Tsallis information
measure (see Eq. (\ref{tsallis})  
below) for a general value of the parameter $q$.

Much work in this respect has been performed
recently (165 refereed papers at the time of this writing). NET has been proved
capable of overcoming the inability of conventional Statistical Mechanics 
(for which one has $q=1$) to
address problems in some fields. We may cite, among such NET successful 
applications, those to astrophysical problems \cite{t5,t6}, 
to L\'evy flights \cite{t7}, to turbulence phenomena \cite{t8}, to simulated
annealing \cite{t9}, etc. The interested reader is referred to \cite{t1} for
additional references.  

Within a NET context, if we associate a 
 probability distribution 
$\{ p_i \}$ to  a set of $W$ different microstates $i$
\be \label{nueva}
\sum_i^W p_i = 1,
\ee
 the concomitant Tsallis information
measure reads  \cite{t1}
\be \label{tsallis}
S_q = (q-1)^{-1}\,\sum_i^W \,p_i\,[1  - p_i^{q-1}].
\ee

It is easy to see that, for $q=1$, one regains the conventional
Boltzmann-Gibbs-Shannon logarithmic form \cite{t1}. Otherwise, we are led into
the realm of nonextensivity \cite{t1,t2,t3,t4,t5}. Indeed, 
in order to study the limit $q \rightarrow 1$ we rewrite 
Eq. (\ref{tsallis}) in the fashion

\be
S_q =  \sum_i p_i \, \left\{\frac{ 1- e^{(q-1) \, \ln p_i} }{q-1} \right\},
\label{(roy11)}
\ee

\noindent
and find that for $q \rightarrow 1$  
  
\be
S_1 \equiv \lim_{ q \rightarrow 1 } \, S_q = -k \sum_i p_i \, \ln p_i,
\label{(roy12)}
\ee

\noindent
i.e., for  $q=1$ Tsallis' entropy coincides with the Gibbs-Shannon one.

\subsection{Entropic and Fisher's measures within a NET context}

Within a NET context one identifies the information measure, of course, with
the Tsallis generalized entropy (\ref{tsallis}). 
 It goes without saying that
extension of Fisher's ideas to a NET scenario should be of interest. In this
respect, some preliminary work has been advanced in \cite{miller}. In that
paper, Fisher's information measure $I$ is generalized to a nonextensive
environment (yielding a suitable quantity $I_q$). 

\subsection{The goal of the present communication}

Here we will show that generalization of the most important physical
application of Fisher's ideas, the so-called Frieden-Soffer principle
of Extreme Physical Information \cite{f7} (to a non-extensive setting) yields
intriguing results concerning 
several important equations of classical dynamics. 
 The concomitant analysis will also illuminate
interesting aspects of the different NET statistics that arise as one varies
the all important Tsallis' $q$-parameter.

As a first step in such a direction, and in order to establish a convenient
notation, we begin by refreshing, in Section II: 
 i) the extension of Fisher-Frieden ideas to non-extensive
settings, and  
ii) the Frieden-Soffer Principle concerning
extremization of Physical Information \cite{f7}.
 In
Section III we discuss the application of this Principle in a
non-extensive environment and 
present the rather surprising results of the present Communication
in Section IV. Finally, conclusions are drawn in Section V.

\section{A review of fundamental notions}

\subsection{Fisher's information in a nonextensive setting}
A suitable generalization of (\ref{ifisher}) should read, according to the
second NET-postulate \cite{miller}

\begin{equation}
I_q\,=\,< \,\{ \frac{\frac{df}{dx}}{f(x)} \}^2\,>_q ,  \label{qfisher}
\end{equation}

\noindent
which is our generalized Fisher's information measure for translation
families.  
In \cite{miller} one uses (\ref{qfisher})
 to

\begin{itemize}
\item  generalize the Cramer-Rao bound \cite{f1}.

\item  discuss connections between Tsallis' entropy, on the one hand, and
Fisher's measure, on the other one.

\item  derive a suitably generalized form of the Frieden-Nikolov's bound 
\cite{katz} to the time derivative $\frac{dS_q}{dt}$. This alternative form
connects the entropy increase to $I_q$.
\end{itemize}

\subsection{ The Frieden-Soffer EPI Principle}

The Principle of Extreme Physical Information (EPI) is an overall physical
theory that is able to unify several sub-disciplines of Physics 
\cite{f7,roy}.
In Ref. \cite{f7} FS show that the Lagrangians in Physics arise out of a
mathematical game between an intelligent observer and Nature (that FS
personalize in the appealing figure of a ``demon", reminiscent of the
celebrated Maxwell's one). The game's payoff introduces the EPI variational
principle, which determines simultaneously the Lagrangian {\it and} the
physical ingredients of the concomitant scenario.

FS \cite{f7,roy} envision the following situation: some physical phenomenon is
being investigated so as to gather suitable, pertinent data. Measurements
must be performed. Any measurement of physical parameters appropriate to the
task at hand initiates a relay of information $I$ (Fisher's) from Nature
(the demon) into the data. The observer acquires information, in this
fashion, that is precisely $I$. FS assume that this information is (of
course!) stored within the system (or inside the demon's mind). The demon's
information is called, say, $J$ \cite{f7,roy} .

Assume now that, due to the measuring process, the system is perturbed,
which in turn induces a change $\delta J$ of the demon's mind. It is natural
to ask ourselves how the data information $I$ will be affected. Enters here
FS's EPI: {\it in its relay from the phenomenon to the data no loss of
information should take place}. The ensuing new Conservation Law states that 
$\delta J=\delta I$, or, rephrasing it 
\begin{equation}  \label{epi}
\delta (I\,-\,J)\,=\,0,
\end{equation}
so that, defining an information loss (or action) ${\cal A}$ 
\begin{equation}  \label{action}
{\cal A}\,\equiv \,I\,-\,J,
\end{equation}
EPI asserts that the whole process described above extremizes ${\cal A}$.
FS \cite{f7,roy} conclude that the Lagrangian for a given physical environment
is not just an {\it ad-hoc} construct that yields a suitable differential
equation. It possesses an intrinsic meaning. Its integral represents the
physical information ${\cal A}$ for the physical scenario. On such a basis
some of the most important equations of Physics can be derived 
\cite{f7,roy}.

Within the present context our demon is working in a non-extensive
fashion. An appropriate nonextensive information $I_q$, to be  presently
introduced, is then required,  
 so that the EPI principle would read
\begin{equation}  \label{actionq}
{\cal A}_q\,=\,I_q\,-\,J,
\end{equation}
i.e., using appropriate $q$-generalizations of the intervening
physical quantities. Thus the Frieden-Soffer information transfer game is
played here according to
\begin{equation}  \label{epiq}
\delta (I_q\,-\,J)\,=\,0.
\end{equation}

Information $I_q$ is defined next.

\section{The workings of FS's demon in a non-extensive scenario}

\subsection{Introductory remarks}

One starts here the FS game by using (\ref{epiq}), i.e., by extremizing
Fisher's generalized information \cite{miller} 
(the dot indicates differentiation with respect to $x$) 
\begin{equation}
I_q=\int f(x)^{q-2\ }\ \dot{f}(x)^2\;dx ,  \label{1}
\end{equation}
and considering a physical scenario in which the knowledge of the
information demon \cite{f7} $J$ is 
  
\begin{equation}
J\,=\,\int {\cal \tau}(f)\ dx ,  \label{2}
\end{equation}
 where the ``information density" ${\cal \tau}$ may depend upon $q$. 

As a generalization of the use of probability amplitudes in the 
$q=1$-theory   
 \cite{f7,roy},  define a new
amplitude $v$ obeying 
\begin{equation} f(x)=v(x)^{\frac 2q} .  \label{5}
\end{equation}
Then the Fisher's Information for translation families acquires the appearance
\be
I_q=\frac{4}{q^2}\int \dot{v}^2\;dx.
\ee

The anzat (\ref{5}) acquires the Fisher's form in the special case $q=1$. Our
generalized Lagragian ${\cal L}$ reads (in terms of $v$)
\be
{\cal L=\;}\frac{4}{q^2}\dot{v}^2 ,
\ee

A general equivalence between the Frieden-Soffer EPI Principle in a
non-extensive setting and the Classical Mechanical single-particle Lagragian,
can be established by letting the knowledge of the information demon
\cite{f7} to acquire the form (\ref{2}). In this case the total 
Lagrangian density (that includes the demon's ${\cal \tau}$) is

\begin{equation}
{\cal L}_T=\frac 4{q^2}\;\dot{v}^2-{\cal \tau}(v).
\label{61}
\end{equation}

This mechanical analog is completed with the changes
$$x \rightarrow  t,$$ and we work in units such that $m=1$,
$$statistical \,\,\,\,amplitude\,\,\, v \rightarrow 
classical\,\,\,\,dynamical\,\,\,\, quantity\,\,\,v, $$ so 
that one  computes the total energy of
the particle 
\begin{equation}
{\cal H}=\frac{\partial {\cal L}_T}{\partial \dot{v}}\;\dot{v}-{\cal L}_T,
\label{7}
\end{equation}
and the energy is given by
\begin{equation}
\frac 4{q^2}\;\dot{v}^2+{\cal \tau}(v)=E ,  \label{81}
\end{equation}
which is now the new first integral of the motion. The corresponding
equation of motion is 

\begin{equation}
\ddot{v}+ \frac{q^2}{8}\frac{d {\cal \tau}}{dv}=0 , \label{91}
\end{equation}
which illustrates the fact that many classical dynamical problems 
 can be 
``linked", via EPI,  to 
 a statistical scenario. An effective mass and effective force, that 
adopt the forms (remember we set above $m=1$ so that our effective masses 
become pure numbers)
\begin{equation}
m_{eff}=\frac{8}{q^2},
\end{equation}

\begin{equation}
F_{eff}=- \frac{d {\cal \tau}}{d v},
\end{equation}
can be deduced from (\ref{91}).

In what follows we assume

\be
\n{lambda}
{\cal \tau}(f) \,=\, \lambda f(x),
\ee
with $\lambda$ a normalization constant.  
Notice here an important difference with the scenario devised by Frieden and
Soffer.   
By their aproach, information $J$ (and equivalently quantity $\tau(f)$) is
solved {\it for} through the use  of a basic property of the measured
phenomenon such as unitarity. Depending upon phenomenon, $J$ is often the
expectation value of the kinetic energy. By contrast, in Eq.
(\ref{lambda}) we assume {\it a form for $\tau(f)$}. Its use is then
equivalent to minimizing information $I$ subject to a simple constraint of
normalization. This would be a principle of minimum Fisher information  
\cite{f8}. The EPI formalism has a different interpretation of the same
situation. Its demon now fixes the phenomenological information $J$ as
being merely a statement of normalization. Since any probability
distribution function must   obey normalization, this represents a minimal
amount of information. Hence the demon is being maximally parsimonious in
allowing information into the data. The measurer is thereby presented with
a scenario of maximum ignorance. 
 
 We have 
\begin{equation}
{\cal L}_T={\cal L}-\lambda \;f=\frac 4{q^2}\;\dot{v}^2-\lambda \;v^{\frac 2q},
\label{6}
\end{equation}
and  are now faced  with 
 {\it the Classical Mechanical Lagragian for the motion of a particle of
effective mass $m_{eff}=8/q^2$ in an external potential}  $V(v)=\lambda \;v^{
\frac 2q}$
and such that the effective force is given by 
$F_{eff}=- \frac{2\lambda}{q} v^{\frac{2}{q}-1}$.

With
\begin{equation}
\frac 4{q^2}\;\dot{v}^2+\lambda \;v^{\frac 2q}=E ,  \label{8}
\end{equation}
one  obtains a first integral of the motion.

In the $v,t$  variables, the equation of motion is 

\begin{equation}
\ddot{v}+\frac{\lambda \;q}4v^{\frac{2-q}{q}}=0 ,
\label{9}
\end{equation}
that, if the real parameter $q$ is set equal to unity (Tsallis' tenets
reduce to the orthodox, extensive Boltzmann-Gibbs ones) (\ref{8}) gives the  
first integral of the motion  for a particle subject to the action
of an effective elastic force $F_{eff}=-\lambda v$.

Our first result here can be paraphrased as follows: 
{\it the Frieden-Soffer demon is seen to have the ability of 
translating  a statistical problem into a
dynamical one} with the transformation

\be
\n{fv}
f \rightarrow v \,\,\,(nonlinear),
\ee
and
\be
\n{xt}
x \rightarrow  t \,\,\,(linear).
\ee

Our second result reads: {\it the choice of $q$ 
determines just  what type of dynamical problem is to be
addressed}.  Examples of such problems follow.

\section{Relativistic and mechanical applications}

In this section we stress the connection between the probability distribution
and the cosmological expansion factor, for a closed Friedman-Robertson-Walker
spacetimes. Also, we list a set of simple mechanical problem connected with a
non-extensive statistical setting.

\subsection{Cosmological scenario}

Friedman-Robertson-Walker spacetimes (FRW) are specially important in
providing cosmological models which are in good agreement with observation. In
this section we apply the equations of gravitation to a space which over its
whole extent is completely homogeneous and isotropic. The metric is given by

\be
\n{m}
ds^2=-dt^2+a^2\left[\frac{dr^2}{1-kr^2}+
r^2\left(d\theta^2+sin^2\,\theta\,d\phi^2\right)\right],
\ee

\no where $a/\sqrt{k}$ is the ``radius of curvature" of the space and $k$ is
the curvature of the space. This metric allows for a particular calculational
simplicity, on account of the high degree of symmetry and the single metric
degree of freedom. Also, we are dealing with a geometry not very different
from that of the actual universe, in its later stages.

In order to give a good representation of the scale factor $a$ as a probability
density, we choose an space with positive curvature that means $k>0$. In this
case, another convenient form for the four-dimensional spherical coordinates is
obtained by introducing in place of the coordinates $r$ the angle $\alpha$
according to $r=\frac{1}{\sqrt k}sin\,\alpha$ ($\alpha$ goes between the limits
0 to 2$\pi$). Then

\be
\n{m1}
ds^2=-dt^2+\frac{a^2}{k}\left[d\alpha^2+
sin^2\,\alpha\left(d\theta^2+sin^2\,\theta\,d\phi^2\right)\right].
\ee

The coordinate $\alpha$ determines the distance from the origin, given by
$\alpha a/\sqrt k$. The surface of a sphere in these coordinates equals
$\frac{4\pi a^2}{k}sin^2\,\alpha$. We see that as we move away from the origin,
the surface of the sphers increases, reaching its maximum value $4\pi
a^2/k$ at a 
distance of $\pi a/2\sqrt k$. After that it begins to decrease, 
and it reduces itself  to a
point at the opposite pole of the space, at a distance $\pi a/\sqrt k$, the
largest distance which can in general exist in such a space if we note that the
coordinate $r$ cannot take on values greater than $1/\sqrt k$.

The gravitational action $S[g]$ can be written as

\begin{equation}  \label{s}
S[g]=S_E[g]+\int{\sqrt{-g}{\cal L}_m\,d^4x},
\end{equation}

\noindent where $S_E$ is the classical Einstein gravitational action, ${\cal
L}_m$ is the matter Lagrangian and $g$ denotes the metric determinant.
Normally the classical Einstein gravitational action $S_E$ is

\begin{equation}  \label{sg}
S_E[g]=\frac{1}{16\pi G}\int{\sqrt{-g}R\,d^4x}+(\mbox{surface terms}),
\end{equation}

\noindent where $G$ is the gravitational constant, $R$ the scalar curvature
and the surface terms (st) are to be chosen so as to cancel those arising from
the integration by parts of the second derivatives terms contained in $R$.
Throughout, we use units that entail $c=1$. The variational problem in
Eq.(\ref{s}) leads to Einstein's equation for the classical geometry

\begin{equation}  \label{einstein1}
R_{ik}-\frac 12g_{ik}R=8\pi GT_{ik},
\end{equation}

\noindent where $T_{ik}$ is the energy-momentum tensor of matter and
fields, and $R_{ik}$ is the Ricci tensor with $i,k=1,..,4$.

For geometries of the form Eq.(\ref{m1}) the scalar curvature is given by

\begin{equation}  \label{r}
R=6\left[\frac{\ddot a}{a}+\frac{\dot {a}^2}{a^2}+\frac{k}{a^2}\right],
\end{equation}

\noindent where the dot denotes derivative with respect to $t$. We see that
for these cases the gravitational action  Eq.(\ref{sg}) acquires  the
appearance

\be
\n{s11}
S[g]=\frac{1}{16\pi Gk^{3/2}}\int_{0}^{2\pi}\int_{0}^{\pi}\int_{0}^{\pi}
\int_{t_1}^{t_2}a^3sin^2\alpha\, sin\theta\, R\, d\alpha\,d\theta\,d\phi\,dt
+(\mbox{st}),
\ee

\no inserting the scalar curvature (\ref{r}) into the gravitational action
(\ref{s}), we get

\begin{equation}  \label{action1}
S[g]=\frac{3\pi}{4Gk^{3/2}}\int_{t_1}^{t_2}{\left(-a\dot a^2+ka\right)\,dt}+
\frac{3\pi}{4Gk^{3/2}}\left[a^2\dot a\right]^{t_2}_{t_1}+(\mbox{st}).
\end{equation}

Taking into account that the surface terms (st) can be chosen to cancel
those arising from the integration by parts of the second derivatives terms
contained in (\ref{r}) the last three terms in (\ref{action}) can be
cancelled. Finally

\be
\n{sf}
S[g]=-\frac{3\pi}{4Gk^{3/2}}\int_{t_1}^{t_2}{\left(a\dot a^2-ka\right)\,dt}.
\ee

Below we shall compare the gravitational action (\ref{sf}) with the
generalized action that arises from a minimization of Fisher's generalized
information $I_q$ as constrained by $J$ 
 (\ref{actionq}) and (\ref{epiq}). It is to be mentioned that EPI has
 been used to derive the Einstein field equations of gravitation,
in a manner quite different from the one here discussed, in Ref. 
\cite{fnew}, where the ordinary ($q=1$) $I$ is employed. Here we are not
directly using EPI, but a variational procedure that involves our Fisher
quantity $I_q$.

>From the identification of equations
(\ref{actionq}), (\ref{1}), (\ref{2}), (\ref{lambda}), (\ref{8}) and (\ref{9})
with (\ref{sf}), we obtain

\be
\n{I_3}
S[g]=-\frac{3\pi}{4Gk^{3/2}}{\cal A}_q=
-\frac{3\pi}{4Gk^{3/2}}\left(I_3-J\right),
\ee

\no for

 $$q=3$$


\no with

\be
\n{af}
a=f,   \qquad t=x,     \qquad  k=\lambda,
\ee

\no and

\be
\n{v}
\frac 4{9}\;\dot{v}^2+k\;v^{\frac 23}=E,
\ee

\no where $E>0$ in order to obtain a real probability distribution $f$ and the
effective force $F_{eff}=-2k/3v^{1/3}$ is attractive. 
At this stage we have
reduced the gravitational closed (FRW) problen to a simple analogous
mechanical problen, which in turn, is equivalent to 
that posed by a non-extensive
statistical scenario with $q=3$\footnote{$I_3$ obeys a modified version 
of the Cramer-Rao bound (see Ref. \cite{miller} for details of this bound 
for $q \ne 1$).}. 

The comparison $a(t)=f(x)$ (from (\ref{af})) shows that {\it the probability
distribution can be identified with the radius of curvature of the space by
choosing the cosmological time variable $t$ proportional to the information
parameter $x$, while the curvature of the space $k$ is identified with
$\lambda$, the normalization constant}. In this gravitational problem the
variable $v=a^{3/2}$ is a function of the radius of curvature and Eq.
(\ref{v}) gives the dynamics for the Robertson-Walker metric when the right
hand side of (\ref{v}) is the energy density of the matter.

To see that, let us write now the $00$ component of Einstein equation with no
cosmological constant for the Robertson-Walker metric in the case of a perfect
fluid source, satisfying an equation of state $p=(\sigma-1)\rho $:

\begin{equation}
\label{H}
3H^2+3\frac{k}{a^2}=8\pi G\frac{\rho_0}{a^{3\sigma}},
\end{equation}

\noindent where $H=\dot a/a$ is the expansion rate and $\rho_0$, $\sigma$ are
constants. Identifying the energy $E$ with the constante $8\pi G/3$, Eq.
(\ref{H}) transforms into Eq.(\ref{v}), with $f(x)=a(t)$ and $\sigma=1$. 
Consequently, we have a universe filled with a classical matter source. In
other words, we have shown the equivalence between the equation of motion
for the matter dominated universe and that equation that 
arises in dealing with a $q=3$ non-extensive statistical scenario.

We thus are led to the non-linear differential
equation (\ref{v}) that should yield our ``optimal" probability distribution
$f$. From (\ref{v}) it is easy to show that the probability distribution $%
f=v^{2/3}<E/k$, i.e., it has an upper positive limit and it is bounded. We
conclude that this result has a physical meaning. Indeed, inserting
$f(x_m)=E/k$ into (\ref{v}) , we obtain $\dot f=0$. Thus, at $x_m$ the
probability distribution $f$ has an extremum. On the other hand, inserting
these results into (\ref{91}) we have $\ddot f=-\frac{k^2}{2E}$ (negatively
defined). An $f$-maximum at $x_m$ ensues.
 According to the identifications 
 $x \rightarrow t$ and  $a(t) \rightarrow f(x)$, we conclude that the
 radius of  curvature $a/\sqrt k$ reaches a  maximum at $x=x_m$ 
(i.e., at a finite time $t_m$).  A 
contraction process takes place afterwards.

\subsection{Mechanical applications}

An interesting set of physical mechanical problem can be interpreted in a
non-extensive statistical setting when the transformation given by Eqs.
(\ref{fv}) and (\ref{xt}) is applied. In the follows we list these problems and
add some pertinent comments . Notice that in all cases no special physical
meaning is to be atached to the statistical quantity $f$ (other than 
it minimizes Fisher's information measure with some constraints). 
What does possess a  
physical meaning is the dynamical quantity $v$ that arises
after implementing the transformation (\ref{fv}).  

\begin{itemize}
\item $q=1$

We deal with the harmonic oscillator potential 
$V(v)=\lambda \;v^2$, where  $v$ is the position
coordinate and $2 \lambda$ the  stiffness constant $k$. 

\item $q=2$

We have a constant  field  with a linear potential $V(v)=\lambda v$.

\item $q=-1$

We face the centrifugal potential for a free 
particle in cylindrical coordinates
$V(v)=\lambda v^{-2}$, with $v$ the radial coordinate and $2 \lambda$ the 
  square of the angular momentum.

\item $q=1/2$

Calling $v=\varphi $ we have $V(v)=\lambda
\varphi ^4$, i.e., Higgs' potential for a massless scalar field, a 
 crucial ingredient (in particle physics) for 
understanding spontaneous symmetry breaking.

\item  $q=-2$ 

We deal with a particle subjected to 
Coulomb's potential  in spherical coordinates: $V(v)=\lambda
v^{-1}$, with $v$ the radial coordinate and $\lambda$ a constant proportional
to the electrical charge of the particle.

\end{itemize}

\section{ Conclusions}

In the present communication we translated the rules of the Frieden-Soffer
Information Transfer game into a NET parlance. By suitably playing the
game we find,   
  exact analytical 
 solutions to special instances of classical dynamics. Two transformations
\begin{itemize}
\item from probability distribution $f$ to dynamical variable $v$ (nonlinear)
\item from space $x$ to time $t$ (linear)
\end{itemize}
{\it allow the demon to convert statistical into dynamical equations}. 
In an orthodox $q=1$ context, he needs to know the expectation value of the
kinetic energy. In a variable $q$ setting, he just needs to ascertain the 
normalization of the probability distribution.

Thus, dynamical solutions can be regarded as probability distributions, in a
variable $q$ NET-environment, that extremize Fisher's information measure for
translation families.  These
solutions can then be added to the impressive collection elaborated by Frieden
and Soffer \cite{f7}.

That a NET environment may be the appropriate one in order to address
simple dynamical  problems  seems to vindicate similar connections already
reported, in a totally different (non-Fisher) context by 
other authors \cite{tsallis,tsallis1}. The present results should be
viewed within such a background-landscape.

The extension to a non-extensive setting of the Frieden-Soffer principle of 
Extreme Physical Information (EPI) is seen here to lead to a rather 
intriguing  
result:     The Frieden-Soffer demon can translate {\it dynamical frameworks}
into {\it statistical ones} and vice versa. Two apparently disparate scenarios
(deterministic and stochastic) are seen to emerge from the {\it same}
Principle. This confirms results obtained in a quite different manner by 
Cocke and Frieden using the standard $q=1$ scenario \cite{fnew}. EPI
($q=1$) can also derive both stochastic and deterministic laws, according
to the nature of the phenomenon under study 
\cite{roy}.  The NET setting adds the 
feature of a {\it direct} translation between deterministic and stochastic
 environments. 
Thus, we believe to have developed here an interesting generalization to the
work of \cite{f7}.

\acknowledgments

The authors are indebted to the PROTEM Program of the National Research Council
(CONICET) of Argentina for financial support. One of us (L.P.C) is indebted to
the University of Buenos Aires for financial support under Grant EX-260. F.P.
also thanks the Fellowship Program of the National University La Plata.


\end{document}